\documentstyle[12pt]{article}
\begin{document}
\voffset-0.5cm
\textwidth=16cm
\textheight=22cm
\def\lb{\\ }
\def\alphamz{$\alpha_s(M_{Z})$}
\def\alphamza{\alpha_s(M_{Z})}
\def\alphamt{$\alpha_s(m_{t})$}
\def\gamud#1#2{$\Gamma^{#1}_{#2}$}
\def\delud#1#2{$\delta^{#1}_{#2}$}
\def\Gf{{\rm G}_F}
%
\begin{titlepage}
\pagestyle{empty}
\begin{flushright}
\vbox{
{\bf TTP 93-4}\\
{\rm February 1993}\\
 }
\end{flushright}
\vspace{0.8cm}
\begin{center}
{\Large\bf The Top Width\large\footnote{Work partly supported by
the BMFT Contract 055KA94P,
and by the Polish Committee for Scientific Research
(KBN) Grants No. 203809101 and 223729102.}
\\
\vskip0.3cm
\large\bf Theoretical update }
\end{center}
\vspace{0.5cm}
\begin{center}
{\bf\Large Marek Je\.zabek\large
\footnote{E-mail: jezabek@chopin.ifj.edu.pl;
Alexander von Humboldt Foundation Fellow}  }\\
    \vskip0.2cm
    {\it  Institute of Nuclear Physics, Kawiory 26a,
    PL-30055 Cracow, Poland}
    \vskip0.3cm
    {\bf\large and}
    \vskip0.3cm
{\bf\Large Johann H. K\"uhn}\\
    \vskip0.2cm
   {\it Institut f\"ur Theoretische Teilchenphysik,}
   {\it Universit\"at Karlsruhe}\\
   {\it Kaiserstr. 12, Postfach 6980,}
   {\it 7500 Karlsruhe 1, Germany}
\end{center}
\vskip1.0cm
\begin{center}
{\bf Abstract}
\end{center}
\begin{quote}
{\rm
A critical assessment of the available calculations of the
top quark width is presented.
QCD corrections, the finite mass of the $b$ quark and the effect
of the $W$ width are included as well as the electroweak
corrections. The relative importance of these corrections
is demonstrated for the realistic range of top masses.
For the QCD corrected decay rate
we use the formulae from \cite{JK1}
and include the electroweak
correction taken from \cite{DS}.  Our results differ from those
available in the literature because all the later calculations
ignored the effect of W width discussed earlier in \cite{JK1}~.
This leads to an effect comparable in size to the electroweak
correction.}
\end{quote}
\end{titlepage}
\pagestyle{plain}
\section{Introduction and summary}
The discovery of the top quark has been anticipated
since many years at accelerators of increasing energy.
Present hopes are based on analyses of high precision
data and the standard theory, see \cite{Rolandi}.
The top is the first heavy quark whose mass can
be measured to better than 1\% precision at a future $e^+e^-$
collider. Therefore, measurements of its width will not only test
the standard model at the Born level, but
also the QCD radiative corrections
which are of order 10\% \cite{JK1}~.
This is in contrast to
$b$ and $c$ quarks, where uncertainties in
the masses and non-perturbative effects preclude
this possibility.

Recently, the complete one loop electroweak corrections to
the total rate have been also calculated \cite{DS,Gad}~,
and turned out to be rather small (1-2\%)~. Nevertheless, it
has been claimed \cite{DS,Gad} that a precise measurement of
the top width may serve as a consistency check for the
electroweak sector of the standard model.
In fact a number of calculations have been performed studying
electroweak effects on the top width in theories extending
the standard model \cite{GH}~. In particular it has been
found that the additional corrections
from the extended Higgs sector of the minimal supersymmetric
standard model are significantly smaller than 1\%.

In this article we give the standard model predictions for the
top quark width. Our results are different form those in
\cite{DS,Gad} because we include the effect of $W$ boson
width considered in \cite{JK1}  and neglected in later works.
This effect is comparable in size to the electroweak corrections.
A number of intrinsic uncertainties remains. The present
uncertainty in $\alpha_s$ and the ignorance concerning the QCD
correction of order ${\cal O}({\alpha_s}^2)$
limit the accuracy of the
prediction to about 1-2\%.
One has to take into account also the errors,
both experimental and theoretical,
in the determination of the top mass.

At present the best place for a precise determination of
$\Gamma_t$ is believed to be the threshold region for
$t\bar t$ production in $e^+e^-$ annihilation. The most
optimistic current estimate of the relative precision is
5\% \cite{Fujii}~.
Therefore, it is mandatory to give the theory prediction
which as the one presented in this article
is accurate up to order of 1\% .

\section{QCD corrected decay rate}
We assume throughout three
families of quarks.  Thus the effects of CKM mixing are
negligible.  The QCD corrected width of the top quark is given by
the following formula \cite{JK1}:
\begin{eqnarray}
\Gamma^{(1)} = {{\Gf}^2 {m_t}^5\over 192\pi^3}
\left( 9 + 6{\alpha_s\over\pi}\right)
\int^{(1-\epsilon)^2}_0 {{\rm d}y\over (1-y/\bar y)^2+\gamma^2}
\left[ {\rm F}_0(y,\epsilon) - {2\alpha_s\over 3\pi}
{\rm F}_1(y,\epsilon)
\right]
\label{eq:1}
\end{eqnarray}
where
$$\bar y= \left( M_W/m_t\right)^2\ ,\qquad\epsilon= m_b/m_t\ ,
\qquad\gamma=\Gamma_{W}/M_W$$
and
\begin{equation}
\Gamma_{W}= {\Gf {M_W}^3\over 6\sqrt{2}\pi}
\left( 9 + 6 {\alpha_s\over\pi}\right)
\label{eq:2}
\end{equation}
The functions
${\rm F}_0(y,\epsilon)$
and
${\rm F}_1(y,\epsilon)$
read \footnote{We slightly simplify an original formula from [1]
using relations between dilogarithms.}
\def\Alambd{\lambda(1,y,\epsilon^2)}
\def\SAL{\sqrt{\Alambd}}
\def\Cyeps{{\cal C}_0(y,\epsilon)}
\def\DILOG{{\rm Li_2}\,}
\def\UW{u_w}
\def\UQ{u_q}
\begin{equation}
{\rm F}_0(y,\epsilon) =
{1\over 2}\SAL\,\Cyeps
\label{eq:3}
\end{equation}
where
\begin{equation}
\lambda(u,v,w) = u^2+v^2+w^2- 2(uv+vw+wu)
\label{eq:4}
\end{equation}
\begin{equation}
\Cyeps =
4[(1-\epsilon^2)^2+y(1+\epsilon^2)-2y^2] \quad,
\label{eq:5}
\end{equation}
and
\begin{eqnarray}
{\rm F}_1(y,\epsilon)=
 \frac{1}{2}\Cyeps(1+\epsilon^2-y)
 \left[ 2\pi^2/3 +4\DILOG(\UW)
     -4\DILOG(\UQ) \right.
\nonumber\\
\left. -4\DILOG(\UQ\UW) -4\ln\UQ\ln(1-\UQ)-2\ln\UW\ln\UQ+\ln{y}\ln\UQ
     +2\ln\epsilon\ln\UW  \right]
\nonumber\\
-2{\rm F}_0(y,\epsilon)
\left[      \ln{y}+3\ln\epsilon-2\ln\Alambd \right]
\nonumber\\
+4(1-\epsilon^2)\left[ (1-\epsilon^2)^2
     +y(1+\epsilon^2)-4y^2 \right]  \ln\UW
\nonumber\\
     +\left[ 3-\epsilon^2+11\epsilon^4-\epsilon^6+
     y(6-12\epsilon^2+2\epsilon^4) -
      y^2(21+5\epsilon^2)+12y^3 \right]\ln\UQ
\nonumber\\
     + 6\SAL(1-\epsilon^2)(1+\epsilon^2-y)\ln\epsilon
\nonumber\\
     +\SAL\left[ -5+22\epsilon^2
     -5\epsilon^4- 9y(1+\epsilon^2)+6y^2\right]
\nonumber\\
\label{eq:6}
\end{eqnarray}
where
\begin{equation}
\UQ= {1+ \epsilon^2 -y -\SAL\over 1+ \epsilon^2 -y +\SAL}
\label{eq:7}
\end{equation}
\begin{equation}
\UW= {1- \epsilon^2 +y -\SAL\over 1- \epsilon^2 +y +\SAL}
\label{eq:8}
\end{equation}

Above threshold for real W production the rate (1)
can be approximated by:
\begin{equation}
\Gamma^{(1)}_{nw} = {{\Gf} {m_t}^3\over 16\sqrt{2}\pi}
\left[ {\rm F}_0(\bar y,\epsilon) - {2\alpha_s\over 3\pi}
{\rm F}_1(\bar y,\epsilon)
\right]\quad,
\label{eq:9}
\end{equation}
a result valid in the narrow width approximation.

Neglecting $\epsilon$ one arrives at the
following relatively compact
expressions:
\begin{equation}
{\rm F}_0(y,0) = 2(1-y)^2 (1+2y)
\label{eq:10}
\end{equation}
and\footnote{This form clearly exhibits limiting
behavior
$$ f(y) = {2\pi^2\over3} -{5\over2} -3y(1+y\ln y)+\dots $$
for small y, and
$$ f(y) = 3\ln(1-y) +{4\pi^2\over 3}-{9\over 2}+\dots $$
for $y\to 1^-$~.
Although stated in the
text, these limits are not manifest in the original formula given
in \cite{JK1}.}:
\begin{eqnarray}
f(y) = {\rm F}_1(y,0)/{\rm F}_0(y,0) =
     {2\pi^2\over3}- {5\over 2}+2\ln y\,\ln(1-y)+4\DILOG y -2y +
\nonumber\\
{1\over1+2y}  \left[ (5+4y)\ln(1-y) +{2y\ln y\over1-y}
     -{4y^3(1-y+\ln y)\over(1-y)^2 } \right]
\nonumber\\
\label{eq:11}
\end{eqnarray}

The formula (1) has been derived in \cite{JK1} and tested in
\cite{JK3,JK4}~.
When applied to charm decays, i.e. in the four fermion limit, it
reproduces the numerical results for the total rate \cite{CM}~.

The formulae (3-6) including the $b$ quark mass corrections have
been tested by a numerical calculation in \cite{JK3}.  Although
performed by the same authors this calculation should be
considered an independent one since it was based on a completely
different technique and matrix elements equivalent to those
derived in the classic papers on muon decays \cite{muon}
in a form adopted in
\cite{AP} for charm decays.
Furthermore we have observed that these formulae after an
appropriate analytical continuation are equivalent to formulae in
\cite{CGN}
describing vacuum polarization effects from heavy quarks in
the W boson propagator.  \\
Independent calculations including
non-zero $b$ quark mass have been performed in [2] and \cite{Gad}.
The authors found a numerical agreement of their results with the
formulae (3-6).

The massless limit, eqs. (10-11), derived in [1] was rederived
and confirmed by a number of groups \cite{Czarnecki}-\cite{LY}.

We proceed now to the discussion of the numerical predictions
for the decay rate and the quality of different approximations.
As our input we use:\lb
$M_W = 80.10$ GeV \cite{Rolandi},
$m_b = 4.7$ GeV,
$\alphamza = .118 \pm .007$ \cite{Altarelli}
and $M_Z = 91.187$ GeV \cite{Rolandi}.\lb
Then \alphamt{} is derived from the formula
\begin{eqnarray}
&\alpha_s(Q) = {4\pi\over b_0 \ln Q^2/\Lambda^2}
\left[ 1 - {b_1\over {b_0}^2}
{\ln\ln Q^2/\Lambda^2\over \ln Q^2/\Lambda^2} \right] \\
\label{eq:12}
& b_0 = 11 - {2\over 3}N_f , \qquad
 b_1 = 102 - {38\over 3}N_f  \nonumber
\end{eqnarray}
for $N_f$=5 quark flavours.
Uncertainties in the input value of \alphamz{} as well as
the second order corrections ${\cal O}({\alpha_s}^2)$~,
which have not been calculated
yet, lead to an error which we estimate to be of order 1\%.
In Table 1 we give our results for the widths obtained from
different approximations as well as from the formula (1).  Since
most other authors present their results in comparison with the
zeroth-order result \gamud{(0)}{nw}  obtained in the narrow width
approximation, we define
\begin{equation}
\delta^{(i)} = \Gamma^{(i)}/\Gamma^{(0)}_{nw} - 1
\label{eq:13}
\end{equation}
where $i = 0,1$
corresponds to the Born and the QCD corrected rate
respectively, and the widths in the numerators include the
effects of the W propagator, cf. eq. (1).  Analogously we define
\delud{(1)}{nw}
which is given by the ratio of the QCD corrected and the Born
widths, both evaluated in the narrow width approximation, and
\delud{(1)}{nw}$(0)$
for massless $b$ quark.

\begin{table}[h]
\begin{tabular}{|r|c|c|c|c|c|c|c|c|c|} \hline
$m_t\ $  & \alphamt & \gamud{(0)}{nw} &\delud{(0)}{}&
\delud{(1)}{nw}$(0)$ &  \delud{(1)}{nw} &
\delud{(1)}{}&  \gamud{(1)}{} & \delud{}{ew}&  \gamud{}{t} \\
{\scriptsize(GeV)} &  & {\scriptsize(GeV)} & {\scriptsize(\%)} &
{\scriptsize(\%)} & {\scriptsize(\%)} & {\scriptsize(\%)} &
{\scriptsize(GeV)} & {\scriptsize(\%)} & {\scriptsize(GeV)} \\
\hline
  90.0& .118& .0234& 11.69 & 7.88 &-3.81&  6.56 &.0249& 0.81& .0251\\
 100.0& .116& .0931&  0.16 &-4.56 &-6.91& -6.89 &.0867& 1.04& .0876\\
 110.0& .115& .1955& -1.44 &-6.81 &-7.83& -9.22 &.1775& 1.20& .1796\\
 120.0& .113& .3265& -1.78 &-7.61 &-8.20& -9.89 &.2942& 1.33& .2982\\
 130.0& .112& .4849& -1.82 &-7.97 &-8.37&-10.08 &.4360& 1.43& .4423\\
 140.0& .111& .6708& -1.77 &-8.15 &-8.44&-10.10 &.6031& 1.51& .6122\\
 150.0& .110& .8852& -1.69 &-8.25 &-8.47&-10.05 &.7962& 1.57& .8087\\
 160.0& .109& 1.130& -1.60 &-8.31 &-8.49& -9.99 &1.017& 1.62& 1.033\\
 170.0& .108& 1.405& -1.52 &-8.34 &-8.49& -9.91 &1.266& 1.67& 1.287\\
 180.0& .107& 1.714& -1.45 &-8.35 &-8.48& -9.84 &1.546& 1.70& 1.572\\
 190.0& .106& 2.059& -1.39 &-8.36 &-8.47& -9.77 &1.857& 1.73& 1.890\\
 200.0& .106& 2.440& -1.33 &-8.36 &-8.46& -9.70 &2.203& 1.76& 2.242\\
\hline
\end{tabular}
\caption{Top width as a function of top mass and the comparison of
the different approximations.}
\end{table}

\section{Electroweak corrections}
The complete one loop electroweak
correction to the standard model top decay have been calculated
in [2] and \cite{Gad}.  If the lowest order width is parametrized by
$\Gf$
and $M_W$,
cf. eqs. (1) and (9), the electroweak corrections are
less than 2\% for realistic top masses.  In particular there are
no sizable effects arising from Yukawa couplings
\cite{IMT}~\footnote{We thank Andre Hoang for
checking that this important result  
is in agreement with [2]
when the latter calculation is restricted to the leading
${\cal O}\left({m_t}^2/{M_W}^2\right)$
contribution \cite{Hoang}.}~.
For $100\ GeV \le\ m_t\ \le\ 200\ GeV$
and Higgs mass $M_H\ \ge\ 100\ GeV$
the potentially large ${\cal O}\left({m_t}^2/{M_W}^2\right)$
contribution from the diagrams with Yukawa couplings are
smaller than 0.2\%~,
and hence much smaller than other,
subleading in $m_t$ terms.
The dependence of the correction on
$M_H$ is weak; see [2] for details.  In the following we assume
$M_H = 100 GeV$~.

Strictly speaking $m_t$, $M_Z$, $M_W$, and $M_H$
cannot be treated
as independent parameters.  The standard model and the existing
data imply a relation between them.
For our choice of the masses one can neglect this
effect, provided $m_t$
is not too close to the present experimental
lower limit.  The corresponding change of the Born width is
-2.6\%, -0.8\%,
and less than 0.3\% for
$m_t =$~90,~100, and~$\ge$~110~$GeV$,
respectively.  Therefore we ignore
the above mentioned relation and treat all the masses as
independent parameters.  If the measured $M_W$
and $M_H$ turned out to
be very different from the values assumed in this paper, it would
be straightforward to evaluate the corresponding change of the
Born width.

The width of the top quark including the electroweak correction
can be evaluated from the formula
\begin{equation}
\Gamma_t = \Gamma^{(1)} \left[ 1 + \delta_{ew} \right]\quad,
\end{equation}
and a simple parametrization
\begin{equation}
 \delta_{ew} (\%) \approx 2 - 1.5\bar y
\end{equation}
has been obtained by us from Table 1 in [2].  The results for
\gamud{}{t}
calculated using (14) and (15) are given in our Table 1.

It should be noted that the size of the electroweak corrections
is comparable to the uncertainties from as yet uncalculated
${\cal O}({\alpha_s}^2)$
corrections and the present uncertainty in the value of
$\alpha_s$.  The
electroweak corrections are furthermore sensitive to the details
of the Higgs sector, as exemplified by the recent calculations in
the context of the two Higgs doublet model \cite{GH}~.

%
\vskip 1cm
{\bf\large\noindent Acknowledgements} \\  \vskip0.1cm
M.J. thanks Lalit Sehgal for a conversation which
stimulated writing this report.
He would like also to acknowledge a research fellowship from
the Alexander-von-Humboldt Foundation which enabled his
stay in the Institut  f\"ur Theoretische Teilchenphysik -
Univ. of Karlsruhe, where a part of this work was done,
and to thank the members of the physics faculty there
for warm hospitality and stimulating atmosphere.
\newpage
%
  \def\PLB #1 #2 #3 {{\it Phys. Lett.} {\bf {#1}B}	(#2)  #3}
  \def\NPB #1 #2 #3 {{\it Nucl. Phys.} {\bf B#1}	(#2)  #3}
  \def\PRD #1 #2 #3 {{\it Phys. Rev.} {\bf D#1}		(#2)  #3}
  \def\PRB #1 #2 #3 {{\it Phys. Rev.} {\bf B#1}		(#2)  #3}
  \def\PR #1 #2 #3 {{\it Phys. Rev.} {\bf #1}		(#2)  #3}
  \def\PP #1 #2 #3 {{\it Phys. Rep.} {\bf#1}		(#2)  #3}
  \def\PRL #1 #2 #3 {{\it Phys. Rev. Lett.} {\bf#1}	(#2)  #3}
  \def\CPC #1 #2 #3 {{\it Comp. Phys. Commun.} {\bf#1}	(#2)  #3}
  \def\ANN #1 #2 #3 {{\it Annals of Phys.} {\bf#1}	(#2)  #3}
  \def\APPB #1 #2 #3 {{\it Acta Phys. Polonica} {\bf B#1}(#2) #3}
  \def\ZPC #1 #2 #3 {{\it Zeit. f. Phys.} {\bf C#1}	(#2)  #3}
  \def\CPC #1 #2 #3 {{\it Comp. Phys. Commun.} {\bf#1}	(#2)  #3}
  \def\SJNP  #1 #2 #3 {{\it Sov. J. Nucl. Phys.} {\bf#1}(#3)  #3}
  \def\YadF  #1 #2 #3 {{\it Yad. Fiz.} {\bf#1}		(#2)  #3}
  \def\IJMPA  #1 #2 #3 {{\it Int. J. Mod. Phys.} {\bf A#1}		(#2)  #3}

\end{document}